\documentclass[]{aa}  

\usepackage{graphicx}

\usepackage{graphicx}
\usepackage{txfonts}
\usepackage{lipsum}
\usepackage{subcaption}         
\usepackage{lscape}             
\usepackage{placeins}           
\usepackage{booktabs}  
\usepackage{orcidlink}
\usepackage{physics}
\usepackage{amsmath}
\usepackage{threeparttable}
\usepackage[inline]{enumitem}

\newcommand{\prottw}{$5.17 \pm 0.05$}
\newcommand{\psitw}{$40\pm4$}

\newcommand{\cosistw}{$0.65^{+0.05}_{-0.04}$}
\newcommand{\incstw}{$49^{+3}_{-4}$}
\newcommand{\gammatw}{$81.5 \pm 1.1$}
\newcommand{\imptw}[1][]{$0.24 ^{+0.12} _{-0.09}$}
\newcommand{\rprjuptw}[1][]{$1.36 \pm 0.012$}
\newcommand{\midprior}[1][]{$\mathcal{U}$}
\newcommand{\perprior}[1][]{$\mathcal{U}$}
\newcommand{\arprior}[1][]{$\mathcal{U}$}
\newcommand{\rpprior}[1][]{$\mathcal{U}$}
\newcommand{\cosiprior}[1][]{$\mathcal{U}$}
\newcommand{\lamprior}[1][]{$\mathcal{U}$}
\newcommand{\vsiniprior}[1][]{$\mathcal{N}$$(14.2,0.5)$}
\newcommand{\zetaprior}[1][]{$\mathcal{N}$$(5.68,0.5)$}
\newcommand{\xiprior}[1][]{$\mathcal{N}$$(1.36,0.5)$}
\newcommand{\kampprior}[1][]{$\mathcal{U}$}
\newcommand{\sysprior}[1][]{$\mathcal{U}$}
\newcommand{\jitterprior}[1][]{$\mathcal{J}$}
\newcommand{\lconeqsumprior}[1][]{$\mathcal{N}$$(0.5432,0.1)$}
\newcommand{\rvoneqsumprior}[1][]{$\mathcal{N}$$(0.706,0.1)$}
\newcommand{\lconeqdiffprior}[1][]{$\mathcal{F}(-0.074)$}
\newcommand{\rvoneqdiffprior}[1][]{$\mathcal{F}(0.2814)$}
\newcommand{\midtw}[1][]{$2458913.3721 ^{+0.0002} _{-0.0003}$}
\newcommand{\pertw}[1][]{$4.7201844 ^{+0.0000011} _{-0.0000012}$}
\newcommand{\artw}[1][]{$8.28 ^{+0.25} _{-0.12}$}
\newcommand{\rptw}[1][]{$0.0726 ^{+0.0003} _{-0.0004}$}
\newcommand{\cositw}[1][]{$0.029 ^{+0.014} _{-0.013}$}
\newcommand{\lamtw}[1][]{$-5 \pm 10$}
\newcommand{\vsinitw}[1][]{$14.4 \pm 0.5$}
\newcommand{\zetatw}[1][]{$5.7 \pm 0.5$}
\newcommand{\xitw}[1][]{$1.4 \pm 0.5$}
\newcommand{\kamptw}[1][]{$262 \pm 24$}
\newcommand{\systw}[1][]{$6 \pm 3$}
\newcommand{\jittertw}[1][]{$0.006 ^{+0.058} _{-0.006}$}
\newcommand{\lconeqsumtw}[1][]{$0.54 ^{+0.10} _{-0.09}$}
\newcommand{\rvoneqsumtw}[1][]{$0.71 \pm 0.10$}
\newcommand{\vsini}{$14.2 \pm 0.5$}
\newcommand{\teff}{$6274 \pm 97$}
\newcommand{\rstar}{$1.925^{+0.064}_{-0.065}$}
\newcommand{\mstar}{$1.464 ^ {+0.076}_{-0.079}$}
\newcommand{\logg}{$4.034 ^ {+0.032}_{-0.033}$}
\newcommand{\feh}{$0.119^ {+0.078}_{-0.066}$}
\newcommand{\age}{$2.33^ {+0.71}_{-0.56}$}

\newcommand{\prot}{$5.300 \pm 0.159$}
\newcommand{\bininc}{$125^{+18}_{-10}$}
\newcommand{\vmag}{$9.487\pm0.021$}
\newcommand{\orbper}{$4.720219(11)$}
\newcommand{\rplanet}{$1.396^{+0.056}_{-0.054}$}
\newcommand{\mplanet}{$2.37\pm0.24$}
\newcommand{\ar}{$6.98^{+0.24}_{-0.23}$}
\newcommand{\aau}{$0.0626^{+0.0011}_{-0.0012}$}
\newcommand{\rp}{$0.0745^{+0.0014}_{-0.0015}$}
\newcommand{\kamp}{$223^{+22}_{-21}$}
\newcommand{\ecc}{$0.073^{+0.092}_{-0.052}$}
\newcommand{\imp}{$0.50^{+0.09}_{-0.19}$}

\usepackage[]{hyperref}
\usepackage{xcolor}
\hypersetup{
    colorlinks,
    linkcolor={red!50!black},
    citecolor={blue!50!black},
    urlcolor={blue!80!black}
}

\newcommand{\fref}[1]{Figure~\ref{#1}}
\newcommand{\tref}[1]{Table~\ref{#1}}

\defcitealias{Rodriguez2021}{R21}

\begin{document}

   \title{TOI-1333Ab is on a well-aligned orbit}

   \subtitle{An aligned hot Jupiter around an F-type star with a mutually inclined stellar companion}


   \author{E.~Knudstrup\inst{\ref{sac},\ref{see}}\orcidlink{0000-0001-7880-594X}
        \and
        M.~L.~Marcussen\inst{\ref{sac}}\orcidlink{0000-0003-2173-0689}
        \and
        S.~H.~Albrecht\inst{\ref{sac}}\orcidlink{0000-0003-1762-8235}
        \and
        M.~S.~Lundkvist\inst{\ref{sac}}\orcidlink{0000-0002-8661-2571}
        \and
        C.~M.~Persson\inst{\ref{oso}}\orcidlink{0000-0003-1257-5146}
        }

   \institute{
       Stellar Astrophysics Centre, Department of Physics and Astronomy, Aarhus University, Ny Munkegade 120, DK-8000 Aarhus C, Denmark\label{sac} 
       \email{emil@phys.au.dk} \and
        Department of Space, Earth and Environment, Chalmers University of Technology, 412 93, Gothenburg, Sweden\label{see} \and
        Department of Space, Earth and Environment, Chalmers University of Technology,  Onsala Space Observatory, 439 92 Onsala, Sweden \label{oso}
    }

   \date{Received September 30, 20XX}

 
  \abstract
   {
   Spin-orbit obliquity measurements of hot-Jupiter systems constrain giant planet migration and tidal evolution.
   In binary systems, combining stellar obliquities with the orbit–orbit angle ($\gamma$) between the planetary and stellar companion orbits provides further insight into the dynamical influence of stellar companions.
   }
   {
   Here we aim to determine the projected obliquity ($\lambda$) of the hot Jupiter TOI-1333Ab ($P\approx4.72$~d, $M_{\rm p}\approx2.4$~M$_{\rm J}$) and place the system in the context of hot-Jupiter migration and tidal realignment in binary systems.
   } 
   {
   We analysed spectroscopic observations obtained during planetary transit to model the Rossiter-McLaughlin effect and derive the projected obliquity. 
   We combined this measurement with published system parameters and constraints on the wide stellar companion orbit to assess plausible migration scenarios.
   }
   {
   We measure a projected obliquity of $\lambda=$~\lamtw$^\circ$, 
   showing that TOI-1333Ab is well aligned with the stellar spin axis of its F-type host star. 
   The low obliquity and its modest eccentricity ($e=$~\ecc) are consistent with either disc-driven migration or high-eccentricity migration followed by efficient tidal circularisation and realignment. 
   With an effective temperature of \teff~K, the host star lies above the canonical Kraft break where the systems are frequently misaligned. 
   Despite this, we find the system to be well aligned.
   In comparison with other planetary systems in binaries, 
   TOI-1333 occupies a relatively isolated region in projected obliquity-orbit-orbit angle ($\gamma=$~\gammatw$^\circ$) space, 
   making it a valuable system for studying the interplay between migration, tides, and stellar companions.
   }
   {}


   \maketitle

\nolinenumbers

\section{Introduction}

When it comes to gauging the processes at work in forming and shaping exoplanet systems,
measurements of the relative orientation 
between the stellar spin axes and planetary orbital planes offer a powerful diagnostic.
While planets formed within protoplanetary discs are expected to begin on well-aligned and nearly circular orbits, 
subsequent dynamical interactions can significantly alter these configurations \citep[e.g.][]{Biddle2025}. 
Measurements of the projected spin-orbit angle (obliquity, $\lambda$) for hot-Jupiter systems have revealed a striking diversity, 
ranging from well-aligned systems to polar and retrograde orbits (see \citealt{Albrecht2022} for a recent review).
This diversity of configurations
challenges simple migration scenarios 
and points towards a combination of dynamical pathways and tidal evolution \citep[a prime example being the hot-Jupiter progenitor TIC~241249530~b;][]{Gupta2024}.

Stellar multiplicity adds an additional layer of complexity to this picture. 
Wide stellar companions can induce secular perturbations that excite orbital eccentricities and inclinations, 
potentially triggering high-eccentricity (high-$e$) migration \citep[e.g.][]{Wu2003,Fabrycky2007}. 
At the same time, 
tidal interactions between close-in giant planets 
and their host stars may act to circularise orbits and damp stellar obliquities, 
with efficiencies that depend sensitively on stellar structure, orbital separation, and planetary mass. 
Recent observational efforts, aided by large-scale astrometric surveys such as the \textit{Gaia} mission \citep{GaiaCollaboration2016}, 
have enabled the characterisation of stellar companion orbits and orbit–orbit angles in hot-Jupiter systems, 
providing new insights into the relationship between stellar multiplicity, spin-orbit misalignment, and planetary system architecture \citep[e.g.][]{Rice2024}.

Through the detection of the Rossiter-McLaughlin \citep[RM;][]{Rossiter1924,McLaughlin1924} effect,
we present a measurement of the projected obliquity 
of the TOI-1333A hot-Jupiter system discovered by \citet{Rodriguez2021}, 
hereafter \citetalias{Rodriguez2021}.
The 
TOI-1333 system is an S-type binary consisting of the bright \citep[$V_\mathrm{T}=$~\vmag;][]{Hog2000} and evolved primary F-type star,
TOI-1333A, and a K-dwarf companion separated at around 560~AU
\citep{Michel2024}. 
The wide stellar companion, together with its close-in massive planet, 
makes TOI-1333 an especially valuable system for probing the interplay between giant planet migration, stellar multiplicity, and tidal evolution.

\section{Observations}\label{sec:obs}

In our analysis to measure the
projected obliquity,
we combine both photometric and spectroscopic observations,
which we detail below.
To measure the orbit-orbit angle 
we use data from {\it Gaia} data release 3
\citep[DR3;][]{GaiaCollaboration2023}.

\subsection{Photometry}
The system was observed by the Transiting Exoplanet Survey Satellite \citep[TESS;][]{Ricker2015} in Sectors 15, 16, 56, and 76.
\citetalias{Rodriguez2021} confirmed and characterised the planet TOI-1333Ab using data from Sectors 15 and 16 along with ground-based photometry, spectroscopy, and imaging. 
In \tref{tab:litpars} we summarise their findings of this hot-Jupiter system.
We downloaded the TESS photometry using the \texttt{lightkurve} package \citep{lightkurve}. 
Sectors 15 and 16 were observed with a cadence of 30~min, and 20-second cadence data are available from Sectors 56 and 76. However, for the latter two sectors we opted for 2-minute cadence data, 
as both the transit duration was long enough and the number of transits large enough 
to well sample the transit and describe its shape well.
Opting for 2-minute cadence data has the advantage 
of limiting computation time when calculating light curves. Additionally, a `blue noise' component has been reported in
the 20~s cadence data \citep{Kalman2025}.

\begin{figure}
    \centering
    \includegraphics[width=\columnwidth]{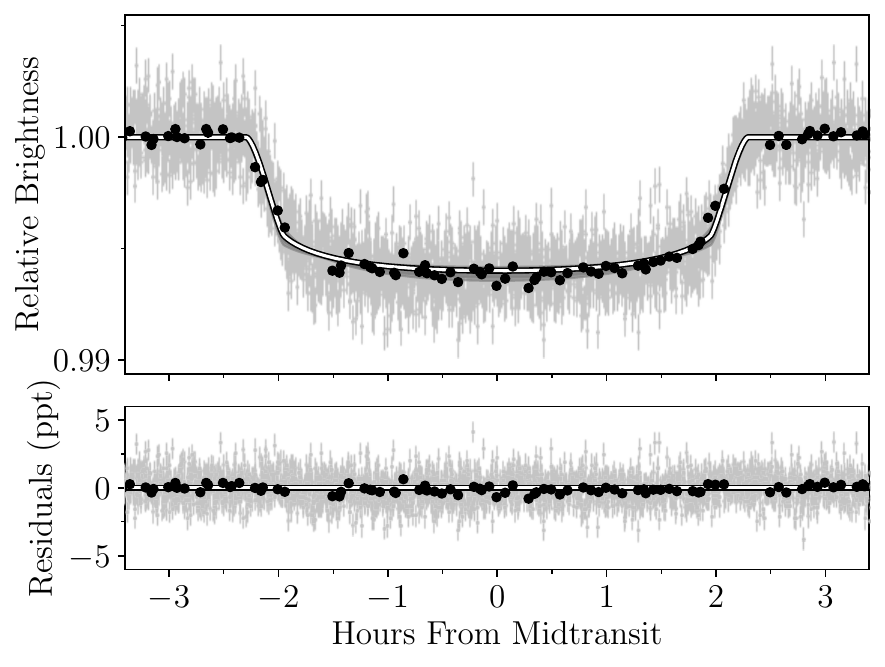}
    \caption{TESS light curve of TOI-1333 folded on the transit of TOI-1333Ab. The black dots demarcate observations taken with a cadence of 30~min. The grey points indicate the 2~min. cadence data. The best-fitting transit model is shown as the solid white line, and the residuals (in parts per thousand) are shown in the bottom.}
    \label{fig:lc}
\end{figure}

After downloading the data, we detrended the light curves 
using Gaussian process (GP) regression \citep{Rasmussen2006}
through the \texttt{celerite} library \citep{celerite}, making use of the Mat\'ern-3/2 kernel. 
Detrending was done in an iterative process (three iterations) consisting of four steps: 
\begin{enumerate*}[label=(\roman*)]
\item We first temporarily removed the transits using the \texttt{batman} package \citep{Kreidberg2015}
starting with parameters from \citetalias{Rodriguez2021}.
\item We then applied the GP to the transit-free light curves fitting the hyperparameters on a sector-to-sector basis.
\item We re-injected the transits
and cut out snippets of the light curve around the mid-transit points, 
including four hours of pre-ingress and post-egress data. 
\item Lastly, we performed a complete analysis (see Section~\ref{sec:anal}) to obtain new parameters.
\end{enumerate*}
The phase-folded light curve is shown in \fref{fig:lc}. In \fref{fig:lc_rot} we show the transit-free light curve along with the GP used for detrending.

In our analysis we chose not to include the
ground-based photometry presented in \citetalias[][]{Rodriguez2021}. Instead, we
only include 
the space-based photometry from TESS.
As the TESS data cover 15 full
and four partial transits,
the four ground-based transits (three separate and some only partial) 
do not add much information.
One benefit of including the ground-based photometry
could be more accurate deblending from contaminant
sources in the TESS apertures,
as pointed out by \citet{Han2025}.
They found that the TESS photometry used in \citetalias[][]{Rodriguez2021}
might have been overcorrected for contamination, resulting in a
planet-to-star radius ratio that is $5.4\pm3.5$\% larger in \citetalias{Rodriguez2021}.
We find a value of \rptw, which is $3\pm3$\% larger than 
-- and therefore also consistent with --
the value reported in \citet{Han2025}. The discrepancy between our value for $R_{\rm p}/R_\star$ and that in \citetalias{Rodriguez2021} might also be a product of the two additional TESS sectors we include.

\subsection{Spectroscopy}
\begin{figure}
    \centering
    \includegraphics[width=\columnwidth]{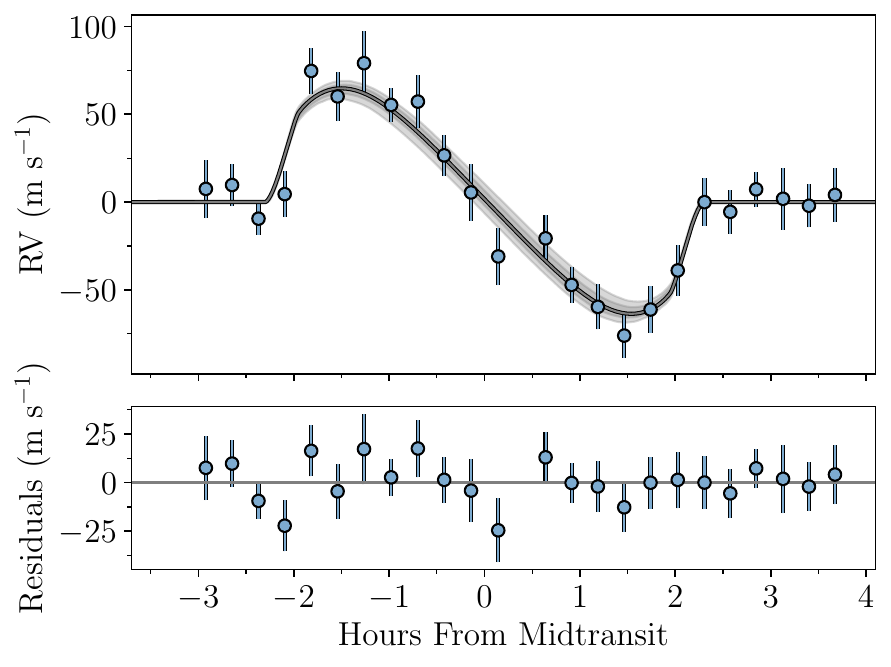}
    \caption{FIES RVs of TOI-1333A after subtracting the best-fitting Keplerian. The solid line denotes the best-fitting model for the RM effect, and the shaded areas represent the 1- and 2-$\sigma$ confidence intervals.}
    \label{fig:rm}
\end{figure}

We observed a transit of TOI-1333Ab on the night of 12-08-2025. 
We made use of the FIES spectrograph \citep{Telting2014}
mounted on the Nordic Optical Telescope \citep[NOT;][]{Djupvik2010}.
We collected 24 exposures--of which 15 were acquired in transit--each with an exposure time of 900~s, 
resulting in a signal-to-noise ratio of around 66 (per pixel at 5,500~\AA).
We extracted the radial velocities (RVs) with a custom implementation of the approach outlined in \citet{Zechmeister2018}. 
The median uncertainty is 13~m~s$^{-1}$ with a root mean square of 10~m~s$^{-1}$. The observations are shown in \fref{fig:rm}, where the RM effect is clearly detected. The FIES RVs are tabulated in \tref{tab:rvs}.

\begin{table}[]
    \centering
    \caption{Literature parameters for the TOI-1333A system.}
    \begin{threeparttable}
        \begin{tabular}{l c}
        \toprule
        Parameter   & Value \\
        \midrule
        Stellar mass,  $M_\star$  (M$_\odot$) & \mstar \\
        Stellar radius,  $R_\star$  (R$_\odot$) & \rstar \\
        Effective temperature, $T_\mathrm{eff}$ (K) & \teff \\
        Surface gravity, $\log g_\star$ (dex) & \logg \\
        Metallicity, $[\mathrm{Fe/H}]$ (dex)  & \feh \\
        Age, $\tau$ (Gyr) & \age \\
        Proj. rot. velocity, $v\sin i_\star$ (km~s$^{-1}$) & \vsini \\
        Binary orbital inclination, $i$ ($^\circ$) & \bininc \\
        \midrule
        Orbital period, $P$ (days) & \orbper \\
        Planet-to-star radius ratio, $R_\mathrm{p}/R_\star$& \rp \\ 
        Scaled semi-major axis, $a/R_\star$ & \ar \\ 
        Semi-major axis, $a$ (AU) & \aau \\
        Velocity semi-amplitude, $K$ (m s$^{-1}$) & \kamp \\ 
        Impact parameter,   $b$ & \imp \\
        Eccentricity, $e$ & \ecc \\
          \midrule
        Planet radius,  $R_\mathrm{P}$ (R$_{\rm J}$) & \rplanet \\
        Planet mass, $M_\mathrm{P}$ (M$_{\rm J}$) & \mplanet \\
         \bottomrule
        \end{tabular}
        \begin{tablenotes}[para,flushleft]
            Selected stellar, orbital, and planetary parameters from \citet{Rodriguez2021}.
        \end{tablenotes}
    \end{threeparttable} 
    \label{tab:litpars}
\end{table} 

\section{Data analysis}\label{sec:anal}

We are interested in measuring two angles:
the angle between the stellar spin axis
of TOI-1333A and the orbital axis of TOI-1333Ab (the projected obliquity, $\lambda$), 
and the angle between the planet’s orbit 
and the binary companion’s orbit (the
orbit-orbit angle, $\gamma$). 
In Section~\ref{sec:prot} we discuss the
strong rotational modulation seen in the light curve.
Should the signal originate from TOI-1333A, 
the stellar inclination would be $i_\star=$~\incstw$^\circ$, and the 3D obliquity would be $\psi=$~\psitw$^\circ$.
We are, however, not convinced that TOI-1333A is the source of the rotational modulation,
mainly because of strong blending from 
a nearby bright star.
We therefore do not claim a measurement for $\psi$.

\subsection{Projected obliquity}

Our approach to extract the projected obliquity is very similar to
that outlined in
\citet{Knudstrup2022},
and we summarise it here.
To determine the best-fitting parameters and their uncertainties,
we performed 
Markov chain Monte Carlo (MCMC) sampling 
using the \texttt{emcee} package
\citep{emcee}. 
We modelled the photometric and spectroscopic data jointly 
to properly account for correlations
between parameters such as
the impact parameter ($b$) and $\lambda$.
This also serves to propagate uncertainties
in the mid-transit time ($T_0$) and orbital period ($P$).

To model the light curves, we used the \texttt{batman} package, 
which is based on the equations in \citet{Mandel2002}.
We used a quadratic limb-darkening law
with
coefficients ($q_1,q_2$) 
from the tables by
\citet{Claret2013,Claret2018} for the appropriate filters.
We stepped in the sum of the limb-darkening coefficients ($q_1+q_2$), 
applying a Gaussian prior to this value with a width of $0.1$,
while keeping the difference ($q_1-q_2$) fixed at the tabulated values.
The same approach for limb-darkening was applied for the spectroscopic data.

We modelled the RM effect using the framework by \citet{Hirano2011}.
We included the effects from macro- ($\zeta$) and micro-turbulence ($\xi$)
using the relations by \citet{Doyle2014} and \citet{Bruntt2010}, respectively.
We adopted these values as Gaussian priors with a width of $1$~km~s$^{-1}$.
We also included the instrumental dispersion from FIES,
which at a resolution of $R\sim67,000$ comes out to $\sigma_{\rm PSF} = 1.9$~km~s$^{-1}$.
Thus, in addition to broadening from macro-turbulence, 
the net Gaussian broadening included in our RM model was $\sqrt{\xi^2+\sigma_{\rm PSF}^2}$. 

We applied a uniform prior to the velocity semi-amplitude ($K$)
to account for the fact that the RV slope observed on a given
night could differ \citepalias[from the value from][]{Rodriguez2021} 
due to stellar variability on a timescale of hours.
Given the very modest eccentricity of \ecc\, reported by \citetalias{Rodriguez2021}, we fixed this value to 0.

When running our MCMC, we
initialised 100 walkers
in a `tight Gaussian ball'
around the parameters from \citetalias[][]{Rodriguez2021} -- when available. 
Convergence was assessed by 
computing the ratio between the length of the chains to the autocorrelation time 
as well as
by visually inspecting the chains and correlation plots.
The ratio in our final run came out to more than 300, 
an indication that the MCMC had converged.
In 
\tref{tab:keypars}
we list some of the key parameters from our MCMC,
and in \fref{fig:rm} and \fref{fig:lc}
we show the best-fitting models for the RM effect and the light curve, respectively.
In \tref{tab:fullpost}
we list the posterior values for the full set of parameters 
along with the priors applied. A correlation plot between $\lambda$, $b$, and $v \sin i$ can be found in \fref{fig:corner}. 
Our final result for the projected obliquity is $\lambda=$~\lamtw$^\circ$.

\begin{table}[]
    \centering
    \caption{Key parameters for the TOI-1333A system.}
    \begin{threeparttable}
        \begin{tabular}{l c}
        \toprule
        Parameter   & Value \\
        \midrule
        Parallax, $\varpi$ (mas)\tnote{(a)} & $5.100 \pm 0.013$ \\
        \midrule
        Projected obliquity, $\lambda$ ($^\circ$) & \lamtw \\
         Projected orbit-orbit angle, $\gamma$ ($^\circ$) & \gammatw \\
        \midrule
        Planet-to-star radius ratio, $R_\mathrm{p}/R_\star$& \rptw \\ 
        Scaled semi-major axis, $a/R_\star$ & \artw \\
        Impact parameter,   $b$ & \imptw \\
        Planet radius,  $R_\mathrm{P}$ (R$_{\rm J}$) & \rprjuptw \\    
         \bottomrule
        \end{tabular}
    \end{threeparttable}
    \begin{tablenotes}[para,flushleft]
        \item[(a)] {\it Gaia} DR3
    \end{tablenotes}
    \label{tab:keypars}
\end{table} 

\subsection{Orbit-orbit angle}

As mentioned, TOI-1333A is part of a binary system; therefore, in addition to the spin-orbit angle we derived the orbit-orbit angle, $\gamma$, which is the minimum mutual inclination between the planetary and stellar companion orbits. It is defined as

\begin{equation}
    \cos \gamma = \frac{\vb{r} \cdot \vb{v}}{\lvert \vb{r} \rvert \lvert \vb{v} \rvert } \, ,
\end{equation}
where $\vb{r}\equiv[\Delta \alpha,\Delta \delta]$ and $\vb{v}\equiv[\Delta\mu_\alpha^*,\Delta\mu_\delta]$ are the relative sky-projected position and velocity vectors. 
Values of $\gamma=0^\circ$ or 180$^\circ$ imply 
that the orbital angular momentum vectors of the planetary 
and stellar companion orbits are (anti-)aligned, 
which, for a transiting planet, 
correspond to the stellar companion orbit being viewed close to edge-on. 
In contrast, $\gamma\approx90^\circ$ indicates that the stellar companion orbit
is nearly perpendicular to the planetary orbit 
and therefore viewed close to face-on.

Using {\it Gaia} DR3 astrometry 
and following the approach outlined in Section 2.4 of \citeauthor{Rice2024} (\citeyear{Rice2024}; but see also \citealt{Behmard2022}),
we calculated the orbit-orbit angle. 
This was done by sampling the astrometric parameters 
of the primary and secondary stars 
while accounting for parameter covariances and boundary effects at $\gamma=0^\circ$ and $180^\circ$.
From this, we obtain $\gamma=$~\gammatw$^\circ$, 
indicating that we see the orbit almost completely face-on.
This is consistent with the
analysis from \citetalias{Rodriguez2021}
who find the orbital inclination to be
\bininc$^\circ$,
thereby ruling out an edge-on orbit for the binary.

\section{Discussion}\label{sec:disc}

Our result for the projected obliquity of \lamtw$^\circ$ suggests 
that TOI-1333A is well aligned with respect to the orbit of TOI-1333Ab.  
In the context of giant planet migration scenarios, 
disc migration offers a straightforward explanation 
for the observed $\lambda$, 
as it has traditionally been considered 
a more quiet form of migration that produces 
well-aligned, near-circular orbits
\citep[e.g.][]{Lin1996}. 
This would also explain the 
very modest eccentricity reported by \citetalias{Rodriguez2021}
of \ecc.

Another scenario is that the planet migrated 
via high-$e$ migration
through von Zeipel-Kozai-Lidov \citep[ZKL;][]{Zeipel1910,Kozai1962,Lidov1962} oscillations.
In this process a stellar perturber 
in an orbital plane that is significantly inclined 
with respect to its planet-(primary) star system
can excite the eccentricity of the proto-hot Jupiter 
through secular interactions
\citep[e.g.][]{Wu2003}.
An inclined 1-M$_\odot$ companion at 1000~AU could, for instance, 
cause ZKL oscillations for a Jupiter-mass planet at 5~AU
on a 20~Myr timescale \citep{Fabrycky2007}.
In this context, the stellar companion, TOI-1333B, with a mass of
$M=0.843$~M$_\odot$ and a (sky-projected) separation
of 560~AU \citep{Michel2024} 
might have played a relevant part 
in shaping the architecture of the planetary system,
similar to how the companions in TIC~241249530
and HD~80606 \citep{Naef2001,Hebrard2010} have been
hypothesised to excite the eccentricity 
(and obliquity) in these systems.

The orbital energy from the eccentric 
(and still wide) orbit of the proto-hot Jupiter
could then be removed by tidal dissipation in the planet
\citep{Fabrycky2007}, 
which would both shrink and circularise the orbit.
If tidal dissipation is at work, 
the initial periastron distance would be around 0.03~AU ($\tfrac{1}{2}a$, $a=$~\aau~AU)
with a circularisation timescale comparable to the age (\age~Gyr) of the system \citep{Socrates2012}.

In addition to the eccentricity,
we might also expect 
the (projected) obliquity to have been excited 
dring high-$e$ migration \citep[e.g.][]{Fabrycky2007}.
Just as tides (raised on the planet) could have circularised the orbit, 
tides raised on the star by the planet might have re-aligned it \citep[e.g.][]{Winn2010,Albrecht2012}. 
Tidal alignment depends strongly on orbital separation
as well as planetary mass,
or rather the planet-to-star mass ratio ($q$).
Given the close proximity ($a/R_\star=$~\artw) 
and mass of the planet ($M_{\rm p}=$~\mplanet~M$_{\rm J}$, $q\sim0.0015\sim 1.6 \times \tfrac{{\rm M}_{\rm J}}{{\rm M}_{\odot}}$), 
TOI-1333A is a system where we expect strong tidal interactions to occur,
and we do indeed find it to be a well-aligned system.

However, 
the tidal realignment timescale has also 
been hypothesised to depend strongly on stellar structure
and may be connected 
to the location of the Kraft break--a 
decrease in 
the rotation rate of 
mid-F spectral types seen 
at an effective temperature of $T_{\rm eff}\sim6500$~K \citep{Beyer2024}.
Based on the distribution of the observed projected obliquities,
this boundary has typically been placed in the range 6100~K to 6250~K,
where stars (hosting hot Jupiters) above this threshold are more
frequently found to be misaligned.
With an effective temperature
of \teff~K,
TOI-1333A falls in the 
frequently misaligned regime
but with a low projected obliquity.

A recent study published by \citet{Wang2025}
statistically delineates
the location of the Kraft break, or more accurately,
the transition in $T_{\rm eff}$,
where hot-Jupiter systems are more frequently found to be misaligned.
\citet{Wang2025} found that by dividing the hot-Jupiter systems
(with $\lambda$ measurements) into either single star systems
or binary (or multiple) star systems, the location 
of the transition for single star systems is very close to the actual Kraft break with
$T_{\rm eff}=6510^{+97}_{-127}$~K. 
On the other hand,
the transition occurs at $T_{\rm eff}=6105^{+123}_{-133}$~K 
for binaries,
and thus closer to the canonical value used for obliquities.

\citet{Wang2025} discuss potential biases for 
this observed shift towards lower temperatures for the Kraft break among binary systems;
one is dilution 
from the (typically) cooler companion,
which shifts the $T_{\rm eff}$ of the host star towards lower temperatures.
In the case of TOI-1333, 
such dilution would bias the inferred $T_{\rm eff}$
of the host star towards lower values, 
implying that the true effective temperature of TOI-1333A could be higher 
than the \teff~K reported by \citetalias{Rodriguez2021}.
However, this value resulted from a multi-component fit to the 
spectral energy distribution (SED), 
which explicitly should account for this dilution.
In addition, the SED fit was constrained by a
spectroscopic estimate of $T_{\rm eff} = 6250 \pm 250$~K 
obtained with the SPC pipeline \citep{Buchhave2012} from TRES spectra \citep{Furesz2008}. 
Given the wide projected separation of the companion ($\sim2.8^{\prime\prime}$), 
dilution is not expected to affect the spectroscopic analysis. 
We therefore consider the effective temperature of
TOI-1333A to be firmly in the hot-star regime.
TOI-1333A thus joins a group of well-aligned binary systems with hot(ter) stars, 
again indicating that multiple factors (such as $a/R_\star$ and $M_{\rm p}$)
are at work in sculpting the orbit.

In any case, 
binary companions 
to the host stars might play a central role
both as triggers for 
high-$e$ migration and in the efficiency 
of the subsequent tidal realignment.
When examining the orbit-orbit angle 
for 20 hot Jupiter, multi-star systems
with projected obliquity measurements,
\citet{Behmard2022} found a tendency
for systems with a misaligned host star ($\lambda>30^\circ$)
to preferentially occur in face-on ($40<\gamma<140^\circ$) configurations,
suggesting a misalignment with the companion star as well.
Contrasting this,
\citet{Rice2024} built a larger sample of $\lambda,\gamma$ systems
and found no such correlation between spin-orbit and orbit-orbit misalignment.
Instead they found 
a tendency for the orbit-orbit distribution 
to peak towards alignment rather than being isotropically distributed,
with a small cluster of `fully aligned' systems ($\gamma\sim0,180^\circ$, $\lambda\sim0^\circ$), which they interpret as demonstrating the role of stellar companions in maintaining order in planetary systems.

Recently, \citet{Giacalone2025} measured the projected obliquity of the hot Jupiter, binary system KELT-23A. They find a retrograde orbit ($\lambda=180\pm5^\circ$) 
and calculate the orbit-orbit angle to be $\gamma=60\pm4$.
They created a toy model 
to simulate the $\lambda,\gamma$ distribution for hot-Jupiter systems
in which the orbits of the planetary and stellar companions are initially aligned with the stellar spin axis of the host star.
After migrating inwards (by some unspecified mechanism), 
the planets tend to have either aligned or polar orbits,
with an excess of systems viewed either face-on or edge-on.

\begin{figure}
    \centering
    \includegraphics[width=\columnwidth]{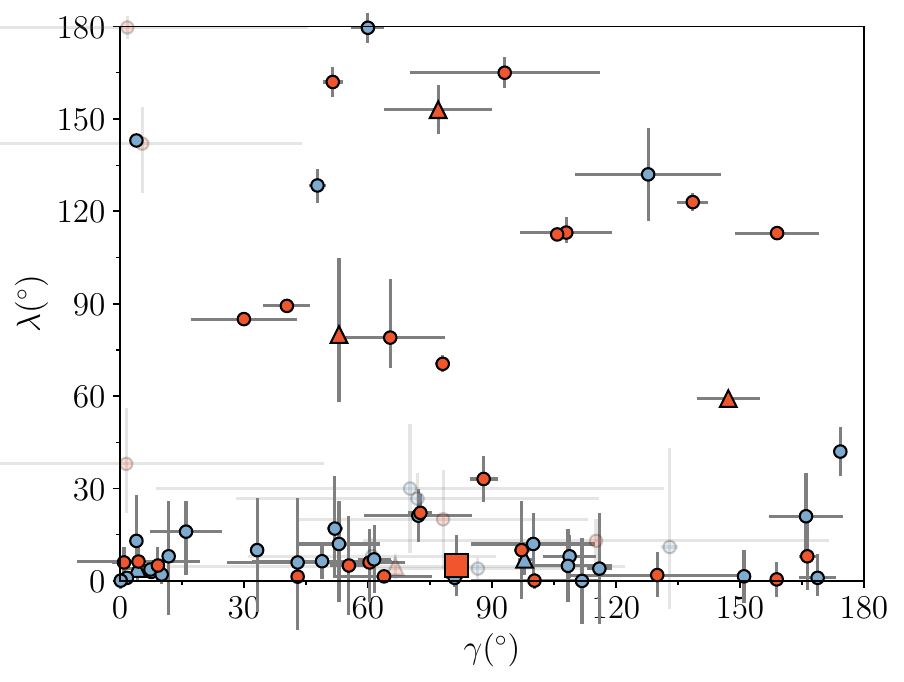}
    \caption{
    Projected orbit-orbit angle ($\gamma$) against the projected obliquity
    ($\lambda$) based on the table by \citet[][]{Rice2024}. 
    Markers are coloured according to the effective temperature of the host star with 
    blue (red) denoting stars cooler (hotter) than $6105$~K, as found to be appropriate for binary systems by \citet{Wang2025}. The triangles demarcate triple star systems, and TOI-1333 is indicated with a square. As in \citet{Rice2024} systems with either $\sigma_\lambda>25^\circ$ or $\sigma_\gamma>25^\circ$ are shown with low opacity.
    }
    \label{fig:lamgam}
\end{figure}

We place TOI-1333 into this context in \fref{fig:lamgam}, 
which shows the projected obliquity plotted against the orbit-orbit angle.
The $\gamma$ values were obtained from the table by \citet{Rice2024}, 
which we have extended with recent measurements both from the literature\footnote{As of 12-01-2026.} and this work (\tref{tab:lamgam}).
From this we find two more of the `fully aligned' systems reported by \citet{Rice2024}\footnote{Although we note they also required the binary inclination to be edge-on, which we do not calculate here.}, namely KELT-3A and WASP-77A. 
At $(\lambda,\gamma)=(5^\circ,85^\circ)$, TOI-1333 sits in a relatively solitary place, with a few more well-aligned systems at $\gamma\sim100^\circ$. 

In the toy model by \citet{Giacalone2025}, stars with $T_{\rm eff}>6250$~K are assigned a lower probability of realignment ($30\%$) after misalignment, resulting in fewer systems at $(\lambda,\gamma)=(0^\circ,90^\circ)$ compared to cooler stars ($80\%$ probability for realignment). 
As was the case in the study by \citet{Wang2025}, 
the TOI-1333 system also appears as an exception 
to the general trend here, making it 
an interesting system for studying the dynamical interactions
that can lead to migration, misalignment, and subsequent realignment.

\section{Conclusions}\label{sec:conc}

Using data from the FIES spectrograph we have measured a projected obliquity of
$\lambda=$~\lamtw$^\circ$
for the hot Jupiter TOI-1333Ab, 
demonstrating that the planetary orbit is well aligned with the stellar spin axis of its host star. 
Such low obliquity is naturally consistent with disc-driven migration, 
which is typically expected to preserve alignment and produce low eccentricities, 
in agreement with the modest eccentricity reported by \citetalias{Rodriguez2021}.
However, the wide stellar companion TOI-1333B also makes high-$e$ migration 
via ZKL oscillations a viable alternative, 
with subsequent tidal dissipation acting to shrink, circularise, and potentially realign the system.
Given the planet's close-in orbit and high planet-to-star mass ratio, 
tidal interactions could have been efficient, 
even though TOI-1333A lies above the canonical Kraft break, 
where hot Jupiter hosts are more frequently found to be misaligned.

In the context of recent studies that suggest 
a lower effective temperature threshold for misalignment in binary systems \citep{Wang2025}, 
TOI-1333A is part of a group of well-aligned exceptions.
This could indicate that tidal realignment may remain effective under a broader range of stellar and dynamical conditions than commonly assumed.

Placed in the wider $\lambda,\gamma$ landscape of planets in binary systems, 
TOI-1333 occupies a relatively isolated region with an orbit–orbit angle of $\gamma=$~\gammatw$^\circ$. 
This combination of low projected obliquity and a near face-on configuration 
underscores the complex role of stellar companions in both exciting and damping misalignments. 
TOI-1333 therefore provides a valuable system for testing competing migration pathways 
and tidal realignment scenarios in hot-Jupiter systems with wide stellar companions.

\begin{acknowledgements}
We are grateful to the anonymous referee for providing insightful comments that helped improve the quality of the manuscript.
This work was supported by a research grant (42101) from VILLUM FONDEN 
and from the Danish Council for Independent Research through grant No.2032-00230B.
M.S.L. is supported by the Independent Research Fund Denmark's Inge Lehmann program (grant agreement No. 1131-00014B).
Based on observations made with the Nordic Optical Telescope, owned in collaboration by the University of Turku and Aarhus University, and operated jointly by Aarhus University, the University of Turku, and the University of Oslo, representing Denmark, Finland and Norway, the University of Iceland and Stockholm University at the Observatorio del Roque de los Muchachos, La Palma, Spain, of the Instituto de Astrofisica de Canarias. The NOT data were obtained under program ID P71-007.
This work presents results from the European Space Agency (ESA) space mission {\it Gaia}. {\it Gaia} data are being processed by the Gaia Data Processing and Analysis Consortium (DPAC). Funding for the DPAC is provided by national institutions, in particular the institutions participating in the Gaia MultiLateral Agreement (MLA). The Gaia mission website is \url{https://www.cosmos.esa.int/gaia}. The Gaia archive website is \url{https://archives.esac.esa.int/gaia}.
This work made use of the following Python packages; NumPy \citep{Harris2020}, SciPy \citep{Virtanen2020}, Matplotlib \citep{Hunter2007}, astroquery \citep{astroquery}, corner \citep{corner}, and Astropy \citep{astropy:2013, astropy:2018, astropy:2022}.
\end{acknowledgements}

%
\bibliographystyle{aa} 
\bibliography{myrefs} 


\begin{appendix}




\section{Stellar inclination and obliquity}\label{sec:prot}

\begin{figure*}
    \centering
    \includegraphics[width=\textwidth]{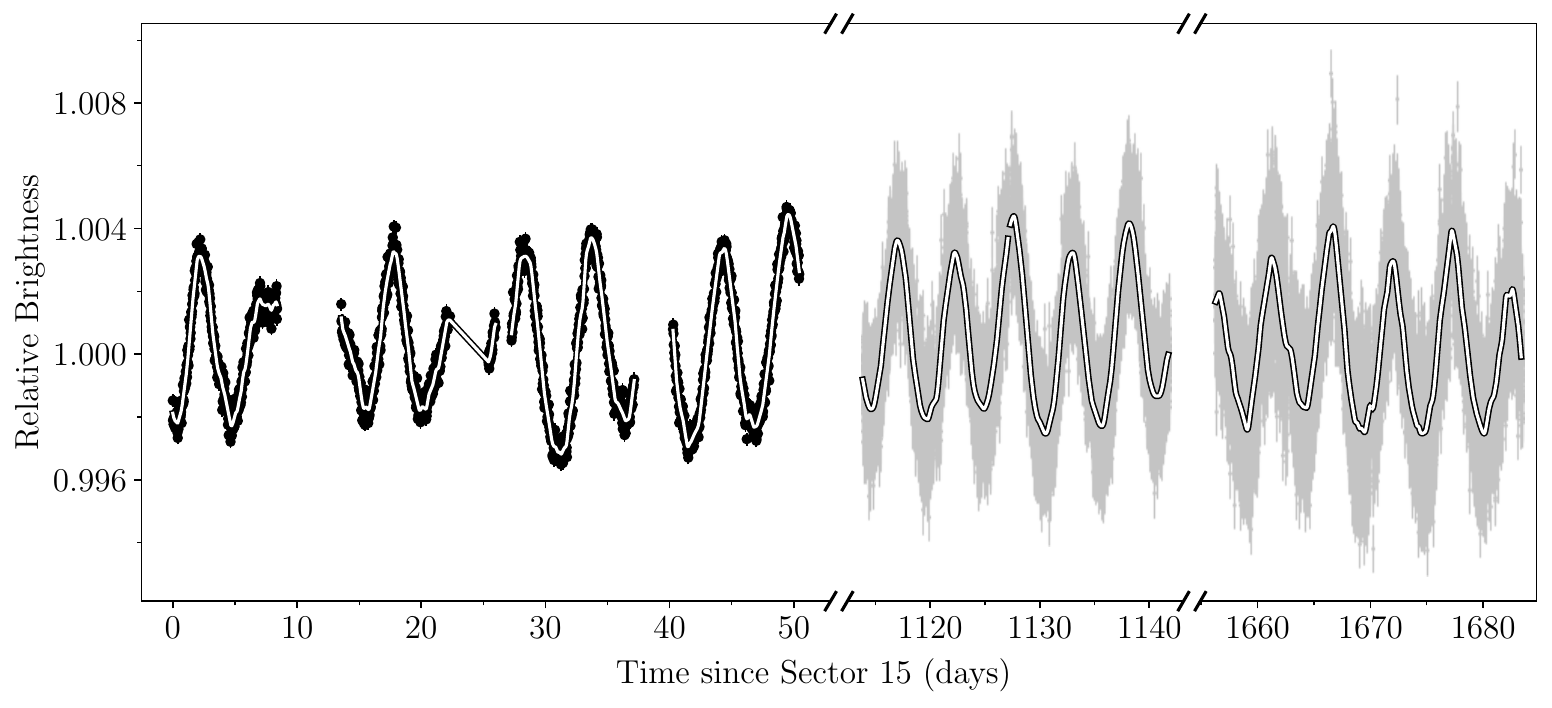}
    \caption{TESS light curves of TOI-1333 showing the rotational modulation, which cannot necessarily be attributed to TOI-1333A. The colour-coding is the same as in \fref{fig:lc}. The GP model used for detrending is shown as the white line with a black outline. Transits were removed by subtracting the best-fitting transit model.
    }
    \label{fig:lc_rot}
\end{figure*}

The light curve from TOI-1333A clearly shows rotational modulation (\fref{fig:lc_rot}).
\citetalias{Rodriguez2021} reported a stellar rotation period of $P_{\rm rot
}=$~\prot~d
based on both the TESS data from Sector 15 and 16 
as well as photometry from the
Kilodegree Extremely Little Telescope \citep[KELT;][]{Pepper2007} survey.
As \citetalias[][]{Rodriguez2021}, 
we note that the rotational modulation seen in the TESS and KELT light curves might be attributed to blended sources instead as both have rather large pixel scales (21$^{\prime \prime}$ and 23$^{\prime \prime}$, respectively).

Should the modulation be coming from a blended source, 
it is most likely a rather bright star in the aperture mask. 
The two brightest stars within the mask are the bound companion, TOI-1333B, 
with $G=12.66$ \citep[][$\Delta G=3.30$ wrt. TOI-1333A]{GaiaCollaboration2023}
and 
TYC 3595-1186-2 with $V=10.15$ and $G=9.98$ ($\Delta G=0.62$ wrt. TOI-1333A). 
TOI-1333B is a K-dwarf \citep[$M_\star=0.843$~M$_\odot$, $T_{\rm eff}=4995$~K;][]{Michel2024}
and therefore seems unlikely to be the source of the modulation given the rather rapid rotation.
It seems more probable that TYC~3595-1186-2 could be the source; 
information on the star is sparse, 
but {\it Gaia} DR2 \citep{GaiaCollaboration2018} gives an effective temperature of $6480$~K and this star therefore seems to be slightly hotter than TOI-1333A.
This is supported by the colour index with $B-V=0.32$ for this star compared to $B-V=0.42$ for TOI-1333A.

What might also suggest that TYC~3595-1186-2 is the source is 
that despite the similarity in brightness, the error on the 
mean flux in the {\it Gaia} $G$-band is around $0.016$\% for TOI-1333A
and $0.033$\% for TYC~3595-1186-2, which could hint at
an additional source of photometric variability for TYC~3595-1186-2.
Spectroscopic follow-up observations of TYC 3595-1186-2 
and of course future resolved photometric time series 
of both TYC 3595-1186-2 and TOI-1333A
could help shed light on the source of the modulation.

Even though we do not believe TOI-1333A 
to be the source of the rotational modulation,
we nonetheless investigated what the stellar inclination
and 3D obliquity would be assuming TOI-1333A is the source.
Following the approach in \citet{Hjorth2021} 
we determined the rotation period using all four TESS sectors
and obtained a value of \prottw~d, consistent with the value from \citetalias{Rodriguez2021}.
As we have now determined $P_{\rm rot}$, we can use this with $v \sin i_\star=$~\vsini~km~s$^{-1}$ and $R_\star=$~\rstar~$R_\odot$ \citepalias[][]{Rodriguez2021} to derive the stellar inclination. We did this using the formulation in \citet{Masuda2020} to account for the fact that $v \sin i_\star$ and $v$ are not statistically independent. From this we found $\cos i_\star = $~\cosistw\, and $i_\star=$~\incstw$^\circ$. 

The (3D) obliquity can then be determined from 
\begin{equation}
    \cos \psi = \cos i_\star \cos i_{\rm o} + \sin i_\star \sin i_{\rm o} \cos \lambda \, ,
\end{equation}
where $i_{\rm o}$ is the orbital inclination. We used our posteriors from Section~\ref{sec:anal} to create the corresponding distributions for $\psi$ resulting in \psitw$^\circ$. 

For context, if $\psi$ were 30$^\circ$ the chance of observing $\lambda$ a factor of two smaller than that is around 35\% \citep[with lower probabilities for larger values of $\psi$;][]{Fabrycky2009}. This also does not favour the value derived for $\psi$ above.
While the arguments presented above are not conclusive, 
we believe the conservative approach is to not claim a measurement for $\psi$.

\section{Additional figures and tables}

\begin{figure}[t!]
    \centering
    \includegraphics[width=\columnwidth]{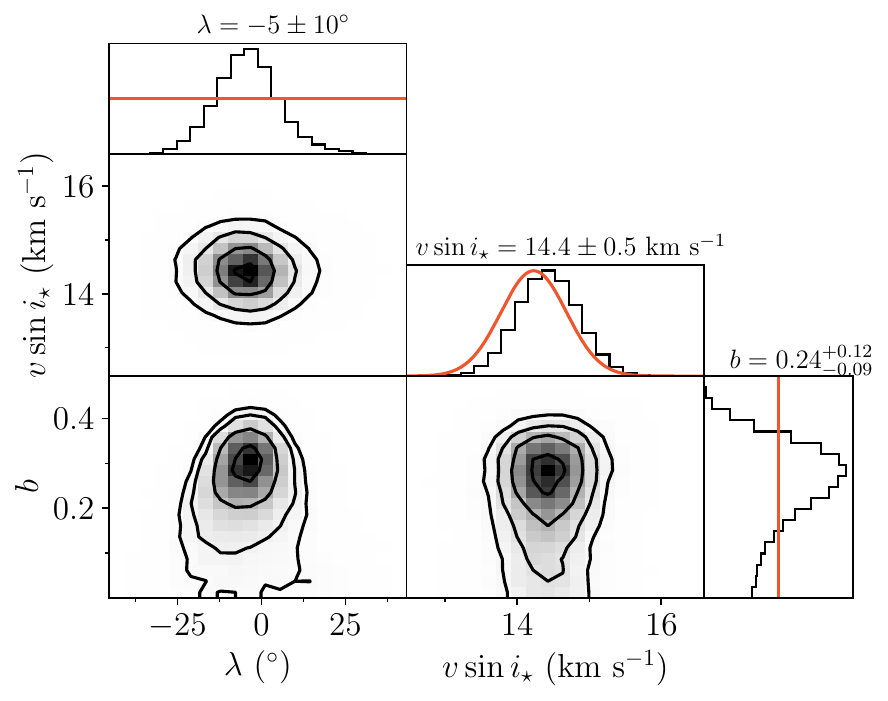}
    \caption{Posterior distributions from our MCMC showing the correlation between $\lambda$, $v\sin i_\star$, and $b$. The red curves in the histogram denote the priors applied.}
    \label{fig:corner}
\end{figure}

\begin{table}[h!]
    \centering
    \caption{FIES RVs of TOI-1333 on the night of 12-08-2025.}
    \begin{tabular}{c c c}
    \toprule
        BJD$_{\rm TDB}$ & RV (m~s$^{-1}$) & $\sigma_{\rm RV}$ (m~s$^{-1}$) \\
    \midrule
        2460900.448148925 & 55.22 & 16.59 \\ 
        2460900.459588338 & 53.52 & 12.03 \\ 
        2460900.471138393 & 30.41 & 9.09 \\ 
        2460900.482575134 & 40.47 & 13.16 \\ 
        2460900.494157147 & 106.64 & 13.12 \\ 
        2460900.505697270 & 88.18 & 14.12 \\ 
        2460900.517229724 & 103.13 & 18.06 \\ 
        2460900.529074778 & 75.28 & 9.57 \\ 
        2460900.540760553 & 73.14 & 14.88 \\ 
        2460900.552259774 & 38.54 & 11.82 \\ 
        2460900.563987611 & 13.37 & 16.06 \\ 
        2460900.575866530 & -27.16 & 16.34 \\ 
        2460900.596544540 & -24.01 & 13.23 \\ 
        2460900.607917608 & -54.43 & 10.10 \\ 
        2460900.619391650 & -70.91 & 12.85 \\ 
        2460900.630892167 & -91.23 & 12.52 \\ 
        2460900.642394491 & -80.35 & 13.35 \\ 
        2460900.654276950 & -62.17 & 14.57 \\ 
        2460900.665964830 & -27.27 & 13.59 \\ 
        2460900.677229890 & -36.65 & 12.57 \\ 
        2460900.688535151 & -27.70 & 10.11 \\ 
        2460900.700278481 & -37.09 & 17.58 \\ 
        2460900.711598352 & -44.89 & 12.37 \\ 
        2460900.722961832 & -42.56 & 15.22 \\ 
    \bottomrule
    \end{tabular}
    \label{tab:rvs}
\end{table}

\begin{table*}[]
    \centering
    \caption{Posterior values for all fitting parameters in our MCMC.}
    \begin{threeparttable}
        \begin{tabular}{l c c c}
        \toprule
            Parameter & Unit & Prior & Value \\
        \midrule
            Orbital period, $P$ & d & \perprior & \pertw \\
            Mid-transit time, $T_0$ & BJD$_{\rm TDB}$ & \midprior & \midtw \\
            Scaled semi-major axis, $a/R_\star$ & & \arprior & \artw \\
            Planet-to-star radius ratio, $R_{\rm p}/R_\star$ & & \rpprior & \rptw \\
            Cosine of orbital inclination, $\cos i_{\rm o}$ & & \cosiprior & \cositw \\
            Projected obliquity, $\lambda$ & $^\circ$ & \lamprior & \lamtw \\
            Projected rotational velocity, $v\sin i_\star$ & km~s$^{-1}$ & \vsiniprior & \vsinitw \\
            Macro-turbulence, $\zeta$ & km~s$^{-1}$ & \zetaprior & \zetatw \\
            Micro-turbulence, $\xi$ & km~s$^{-1}$ & \xiprior & \xitw \\
            Velocity semi-amplitude, $K$ & m~s$^{-1}$ & \kampprior & \kamptw \\
            Systemic velocity, $\Gamma$ & m~s$^{-1}$ & \sysprior & \systw \\
            RV jitter, $\sigma$ & m~s$^{-1}$ & \jitterprior & \jittertw \\
            Sum of LD coefficients TESS, $(q_1+q_2)_{\rm TESS}$ & & \lconeqsumprior & \lconeqsumtw \\
            Difference in LD coefficients TESS, $(q_1-q_2)_{\rm TESS}$ & & \lconeqdiffprior & $\cdots$ \\
            Sum of LD coefficients FIES, $(q_1+q_2)_{\rm FIES}$ & & \rvoneqsumprior & \rvoneqsumtw \\
            Difference in LD coefficients FIES, $(q_1-q_2)_{\rm FIES}$ & & \rvoneqdiffprior & $\cdots$ \\
            Eccentricity, $e$ & & $\mathcal{F}(0)$ & $\cdots$ \\
        \bottomrule
        
        \end{tabular}
        \begin{tablenotes}[para,flushleft]
            \setlength\labelsep{0pt}
            \medskip
            \scriptsize{
                \item Notes: $\mathcal{U}$ ($\mathcal{J}$) denotes a uniform (Jeffrey's) prior. $\mathcal{N}(a,b)$ means that a Gaussian prior was applied with $\mu=a$ and $\sigma=b$. $\mathcal{F}(c)$ means that this value was fixed to $c$.} 
        \end{tablenotes}
    \end{threeparttable}    
    \label{tab:fullpost}
\end{table*}

\begin{table*}[]
    \centering
    \caption{Extension to the table by \citet{Rice2024}.
    }
    \begin{threeparttable}
        \begin{tabular}{l l l c c c}
            \toprule
            Host star & {\it Gaia} DR3, primary & {\it Gaia} DR3, secondary & $\gamma$~($^\circ$) & $\lambda$~($^\circ$) & Source \\
            \midrule
            K2-329A & 2632880927441296640 & 2632880991865749376 & $72\pm45$ & $27^{+9}_{-8}$ & (1) \\
            KELT-3A & 806492023789218688 & 806492023788937216 & $1\pm5$ & $-5\pm4$ & (2) \\
            KELT-4A\tnote{a} & 727624020367528576 & 727624020367030528 & $52.9 \pm 1.2$ & $80^{+25}_{-22}$ & (2) \\
            KELT-23A & 1644692064543192704 & 1644692068838995840 & $60\pm4$ & $180\pm5$ & (3) \\
            TOI-1259A & 2294170838587572736 & 2294170834291960832 & $42\pm17$ & $6^{+21}_{-22}$ & (4) \\
            TOI-1333A & 1978027916667478656 & 1978027912379523712 & \gammatw & \lamtw & (5) \\
            TOI-1759A & 2216420110788943744 & 2216420969782393472 & $116\pm3$ & $-4\pm18$ & (6) \\
            TOI-3714A & 178924390478792320 & 178924390476838784 & $167\pm9$ & $21^{+14}_{-11}$ & (7)  \\
            TOI-5293A & 2640121486388076032 & 2640121482094497024 & $53\pm10$ & $-12^{+19}_{-14}$ & (7) \\
            WASP-77A & 5178405479961698048 & 5178405479961475712 & $11.8\pm0.8$ & $8^{+19}_{-18}$ & (8) \\
            WASP-129A & 5380888758195682944 & 5380888380237386880 & $61\pm28$ & $112\pm1$ & (9) \\
            WASP-140A & 5094154336332482304 & 5094154336332482176 & $81\pm2$ & $-1\pm3$ & (10) \\
            WASP-173A & 2323985539482908416 & 2323985535188372480 & $133\pm2$ & $11^{+32}_{-20}$ & (2) \\
            WASP-193A & 5453063823882876032 & 5453063828179326976 & $52\pm2$ & $17^{+17}_{-16}$ & (11) \\
            \bottomrule         
        \end{tabular}
        \begin{tablenotes}[para,flushleft]
             With the exception of KELT-23A and WASP-129A, the sky-projected orbit-orbit angle has been calculated in this work, while the projected obliquity is from the cited sources:
            (1) \citet{Espinoza2025}; 
            (2) \citet{Knudstrup2024};
            (3) \citet{Giacalone2025};
            (4) \citet{Veldhuis2025};
            (5) This work;
            (6) \citet{Polanski2025};
            (7) \citet{Weisserman2025}
            (8) \citet{Zak2024};
            (9) \citet{Zak2025b};
            (10) \citet{Zak2025};
            (11) \citet{Yee2025}.
            \item[a] Triple star system. 
        \end{tablenotes}
    \end{threeparttable} 
    \label{tab:lamgam}
\end{table*}

\end{appendix}
\end{document}